# Compensation of Multicore Fiber Skew Effects for Radio over Fiber mmWave Antenna Beamforming

Thomas Nikas, Evangelos Pikasis, Sotiris Karabetsos, and Dimitris Syvridis

*Abstract*—In 5G networks, a Radio over Fiber architecture utilizing multicore fibers can be adopted for the transmission of mmwave signals feeding phased array antennas. The mmwave signals undergo phase shifts imposed by optical true time delay networks, to provide squint free beams. Multicore fibers are used to transfer the phase shifted optical signals. However, the inter-core static skew of these fibers, if not compensated, distorts the radiation pattern. We propose an efficient method to compensate the differential delays, without full equalization of the transmission path lengths, reducing the power loss and complexity. Statistical analysis shows that regardless of the skew distribution, the frequency response can be estimated with respect to the rms skew delays. Simulation analysis of the complete Radio over Fiber and RF link validates the method.

## I. Introduction

Forthcoming 5G new radio deployments, adopt a dense Cloud Radio Access Network (C-RAN) architecture, which favors the centralization of the Baseband Units (BBUs) and simplification of the remote radio heads (RRHs), and also mandate for increased bandwidth requirements in mmwave bands with pico- and fempto-cell utilization [1, 2]. Radio over Fiber (RoF) is a promising technology to not only serve network densification and flexible resource allocation, but also to move the mmwave generation at the Central Unit (CU), simplify the RRHs and support the pico/fempto cell network with accurate and stable carrier frequencies [3, 4] as well as antenna beam-forming (A-BF) [5].

In A-BF, the enhanced bandwidth narrow beams impose the use of true time delay (TTD) elements for phase shifting of the currents feeding the antenna to avoid beam squint. Efficient TTD optical beam forming networks (OBFNs) have been proposed to provide multiple squint-free RF beams [6]. OBFNs are installed at the CU since they require thermal stabilization and multiple control signals. There, the modulated optical carriers undergo the required time delays per antenna element and transmitted via optical fiber to the RRHs in which the mmwave signals are generated, amplified and feed the antenna elements. In order to preserve the RF phase relations, equal optical and electrical path lengths are required after the OBFN. Although the electrical path length equality can be easily achieved, since the RF circuitry at RRH consists of photodiodes, transimpedance and mmwave amplifiers, most challenging is the optical path length equality, taking into account that each fiber carries the modulated and phase shifted signal for each antenna element and that the distance between the CU and the RRHs can range to several hundred meters. Hence, path length inequalities produce additional phase shifts. A bundle of single mode fibers (SMFs) can be used, but temperature and stress depended variations lead to unacceptable differential phase errors that distort the antenna array radiation pattern [7]. An efficient solution to this problem is the use of multi-core fibers (MCFs) [7]. MCFs have been proposed for high capacity transmission systems as well as mmwave RoF links [5, 8, 9]. Nevertheless, homogenous MCFs are reported to exhibit inter-core maximum static skew in the order of 0.5 ns/km, while the temperature imposed maximum dynamic skew is 0.06 ps/km/°C [7]. Hence, the major factor affecting the antenna radiation pattern quality is the value of the static inter-core skew. The radiation pattern distortion produced by the MCF depends on the length of the ARoF link and the frequency band of interest. Mmwave bands undergo more severe distortion since phase shifts are proportional to the radio frequency for given skew delays. At 26 GHz for example, 1 ps skew results in 10º phase mismatch. A small phase mismatch may also be encountered due to MCF fan in/fan out modules and fiber coupling. The compensation of the static skew can be performed either in the optical or the electrical domain. Switchable optical delay lines can be inserted at the output of the OBFN or coaxial lines of variable lengths can be inserted at the RRH RF amplifier chain. Both introduce a substantial power loss and the inserted line length must be accurate in a small fraction of the mmwave wavelength. Except from introducing additional power loss and requiring high length

---

Manuscript received XX, 2019. This work was supported in part by the EU project blueSPACE (H2020-ICT2016-2, 762055).

T. Nikas and D. Syvridis are with the Department of Informatics and Telecommunications, National and Kapodistrian University of Athens, Ilissia, GR-15784, Greece (emails: tnikas@di.uoa.gr, dsyvridi@di.uoa.gr).

E. Pikasis is with the Eulambia Advanced Technologies Ltd, Ag. Paraskevi GR-15342, Greece (email: evangelos.pikasis@eulambia.com).

S. Karabetsos is with the department of Electrical and Electronics Engineering, University of West Attika, Egaleo GR-12244, Greece (email: sotoskar@uniwa.gr).

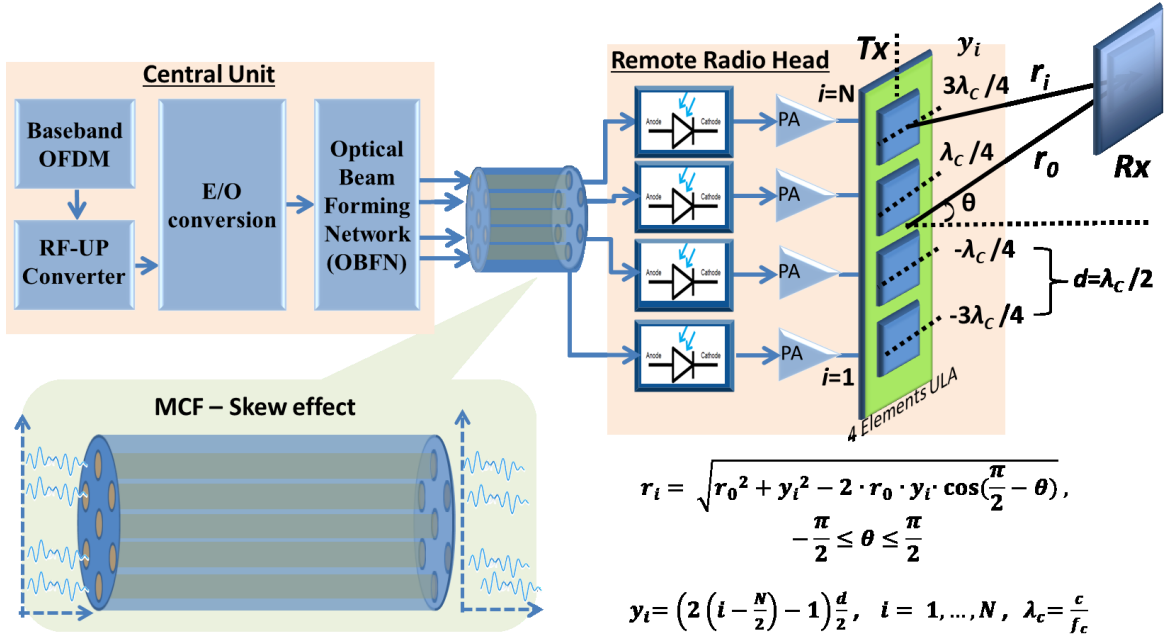

Fig. 1: The RoF link with OBFN and radio propagation model for N=4, d=$\lambda_c$/2.

accuracy, the delay elements either optical or electrical impose power imbalance among the antenna elements.

In this work, we have contacted a theoretical analysis quantifying the MCF static skew effects on the antenna array factor (AF) and the RF channel frequency response as announced in [10]. Moreover, we have also introduced an efficient compensation process without full equalization of the transmission path lengths, reducing the power loss and complexity as well as a statistical analysis of the skew effects. The analysis includes the MCF crosstalk and evaluates its effects. Finally, we have conducted a simulation analysis of a complete RoF and RF link in order to validate the method. The rest of the paper is organized as follows. In sections II and III we present the analysis and partial compensation method respectively, adopting the RoF architecture shown in Fig. 1. In sections IV and V we perform statistical analysis of the proposed method using various distributions of the differential delays, which provides concrete estimations on the expected performance. Section VI presents the numerical analysis of a complete RoF/RF link that validates the obtained theoretical results. Section VII concludes the paper.

## II. THEORETICAL BACKGROUND AND MODEL

In order to evaluate the MCF skew effects on the antenna AF, we consider $N$ decoupled, equally spaced isotropic radiating elements in a uniform linear array (ULA), each element exhibiting flat frequency response within the signal bandwidth. It should be noted that skew delays only affect the AF and not the individual element radiation pattern. The element spacing $d$ is set to a typical value of $\lambda_c/2$ in order to avoid grating lobes when the main lobe is steered towards the limit of $\theta$=90°, with slightly reduced directivity at the broadside ($\theta$=0°) bearing [11]. Nevertheless, our analysis is general, providing the means to calculate the AF and frequency response whatever spacing is used.

The modulation format proposed for 5G is the orthogonal frequency division multiplexing (OFDM) with its variants, so the baseband signal $s(t)$ is expressed as $s(t) = \sum_{k=-\frac{N_F}{2}}^{\frac{N_F}{2}} X_k e^{\frac{j2\pi kt}{T_s}}$, where $X_k$ is the complex QAM symbol modulating the $k$-th subcarrier, $T_s$ is the OFDM symbol duration without the cyclic prefix and $N_F$ is the number of available subcarriers. The baseband signal is up-converted and then modulates the data carrying wavelength. After passing through the OBFN, the optical signals are launched to $N$ cores of an MCF, each core corresponding to an antenna element. After transmission, the inter-core skew imposes differential time delays $\tau_i$, $i$=1…$N$ between each core and the reference core [7]. At RRH, the fast photodiodes produce $N$ photocurrents. Photocurrent amplitude imbalance related to optical power discrepancies may be encountered in the OBFN and MCF - ARoF link which can be compensated using variable gain control at the RRH amplifiers (fig. 1). After bandpass filtering around the carrier frequency $f_c$ and neglecting any non-linearity and noise accumulation, these currents are read as $I(i,t) = I_{0,i} \cdot s(t - \tau_i) \cdot e^{j2\pi f_c t}$, $i = 1,..,N$, where $I_{0,i}$ is the normalized complex amplitude of the $i$-th photocurrent, $I_{0,i} = |I_{0,i}|e^{-j(i-1)\pi \sin\theta_0}$ and $|I_{0,i}|$ is the same for all $i$, if uniform antenna element excitation and no side-lobe suppression measures are considered. The phase delay step $\pi \sin\theta_0$ is imposed by the OBFN to steer the beam to $\theta_0$. The received signal is calculated using the propagation model depicted in fig.1. The photocurrents feed the isotropic transmitting antenna elements. The electrical far field $E_R(i,t)$ of the transmitted wave is proportional to $I(i,t)$ [12, 13], so the field from the $i$-th antenna element at distance $r_i$ from the receiver is $E_R(i,t) = aI(i,t)e^{-j2\pi f_c r_i/c}$, where $a$ is a factor expressing the linear relation of the electrical field and antenna current as well as line of sight path loss. The factor $a$ is assumed time invariant and frequency independent within the signal bandwidth, in order to analyze solely the effects of inter-core skew on the AF.

The receiver antenna is considered isotropic and it is located at distance $r_0$ from the transmitter. The distance $r_0$ is long enough to ensure that the receiver is located at the transmitter far field. The total field $E_R(t)$ at the receiver antenna is the vector sum of all the fields generated from each transmitter antenna element:

$$E_R(t) = \sum_{i=1}^{N} E_R(i,t) = \sum_{i=1}^{N} aI_{0,i}s(t-\tau_i)e^{j2\pi f_c t} e^{-j2\pi f_c r_i/c} \quad (1)$$

The receiver antenna current can be expressed as $b \cdot E_R(t)$ [12]. The factor $b$ expresses the linear electrical field – antenna current relation, assumed to have the same time invariability and frequency independence properties of $a$. The complex baseband received signal $R(t)$ after down-conversion from $f_c$ is $R(t) = Re\{b \cdot E_R(t)\}[e^{j(2\pi f_c t + \pi)}]^* = \sum_{i=1}^{N} ab \cdot I_{0,i}s(t-\tau_i)e^{-j2\pi f_c r_i/c}$, which after analog to digital conversion and FFT is transformed to the frequency domain $R_k = ab \sum_{i=1}^{N} I_{0,i}X_k e^{-j2\pi(f_c+k/T_s)\tau_i} e^{-j2\pi f_c r_i/c}$. The inter-core differential delays $\tau_i$ result in phase delays proportional to the subcarrier frequency. Supposing that the transmitted OFDM symbols are pilot symbols for channel estimation, a zero forcing frequency domain equalizer will output the transfer function:

$$H_k = \frac{R_k}{X_k} = ab \sum_{i=1}^{N} I_{0,i} e^{-j2\pi(f_c+k/T_s)\tau_i} e^{-j2\pi f_c r_i/c} \quad (2)$$

The transmitted QAM symbols $X_k$ are perfectly demodulated at the receiver in the absence of noise. Since $r_i$ is a function of the angle $\theta$, the quantity

$$P(\theta) = \sum_{k=-\frac{N_F}{2}}^{\frac{N_F}{2}} \frac{H_k \cdot H_k^*}{N_{Fu}} = \sum_{k=-\frac{N_F}{2}}^{\frac{N_F}{2}} \frac{|H_k|^2}{N_{Fu}} \quad (3)$$

is the total received power at angle $\theta$, normalized to the number of used subcarriers $N_{Fu}$, $N_{Fu} \leq N_F$. Using (2) and (3) we can calculate the frequency response and the normalized total received power at any $\theta$ and consequently the array factor $P(\theta)$.

The inter-core static and dynamic skew values presented in the literature are not dealing with the statistical properties of the measurements. Therefore, we assume that the random time delays $\tau_i$ are uniformly distributed ranging from 0 to a maximum $\tau_{max}$ value, with $\tau_{max}$ depending on the optical link length. In addition, except from inter-core skew, the MCFs, especially the homogenous ones, demonstrate inter-core crosstalk effects. These effects can be modeled by substituting the $N$ vector values $I_{0,i}e^{-j2\pi(f_c+k/T_s)\tau_i}$ in (2) with $\sum_{j=1}^{N} x_{ij}I_{0,j}e^{-j2\pi(f_c+k/T_s)\tau_j}$, $\forall i$, where $x_{ij}$ are the crosstalk coefficients.

In order to perform comparisons using variable element number $N$, the total transmitted electrical power is normalized:

$$\sum_{i=1}^{N} |I_{0,i}|^2 = 1 \quad (4)$$

In the following section, an MCF Analog RoF (ARoF) - RF beamforming paradigm is applied to the theoretical model, introducing and evaluating the proposed compensation method.

## III. STATIC SKEW PARADIGM AND COMPENSATION

At first, we calculate the transfer function $|H_k|^2$ and the AF $P(\theta)$ for a linear antenna array of $N=4$ elements, at a carrier frequency of 26 GHz. The OFDM signal is composed of 3332 subcarriers with FFT/IFFT length of 4096 and 240 kHz spacing, $T_s = 1/(240\ kHz)$, resulting in a signal bandwidth of 800 MHz. These parameters cover the numerology of 5G specifications and beyond. It should be noted that our analysis is more general in terms of modulation formats and not specifically based on OFDM. On the other hand, OFDM is straightforward in acquiring the channel transfer function at the frequency domain. The antenna elements are fed with uniform current distribution and $|I_{0,i}| = 1/\sqrt{N}$ following the total power constraint in (4). The beam steering angle was set to $\theta_0=0°$ and the inter-core differential delays set to zero. The corresponding $P(\theta)$ and $|H_k|^2$ values at $\theta_0=0°$ are depicted in fig. 2 and fig. 3 respectively (blue lines). $|H_k|^2$ is flat over the entire signal bandwidth and the AF corresponds to a typical linear array of four $\lambda_c/2$ spaced elements.

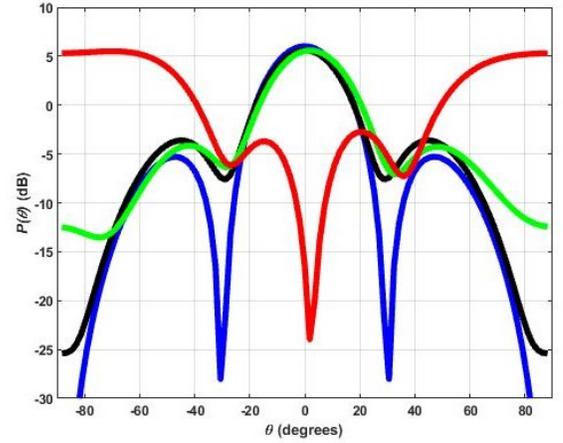

Fig. 2: Calculated array factor. Blue: w/o skew, red: w/ skew, black: w/ compensation, green: w/ compensation and -20 dB crosstalk.

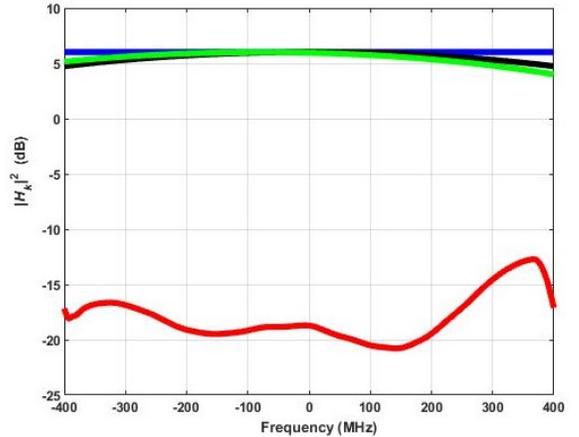

Fig. 3: Calculated frequency response for $\theta_0=0°$. Blue: w/o skew, red: w/ skew, black: w/ compensation, green: w/ compensation and -20 dB crosstalk.

Next, we calculated $|H_k|^2$ at $\theta_0$ and $P(\theta)$ imposing delays with values of 0.00, 0.44, 0.50, 0.10 ns which correspond to 1 km of RoF transmission through cores 2, 3, 6, and 7 of the fiber studied in [7]. Core 2 is considered as reference. $P(\theta)$

and $|H_k|^2$ are depicted in fig.2, fig. 3 with red lines respectively. As expected, $|H_k|^2$ resembles to the transfer function of a frequency selective channel, which in general is non-symmetric. This is attributed to the vector summing of fields produced from antenna elements fed with variable delays, as in the case of multipath propagation. Moreover, the antenna radiation pattern is severely distorted due to the phase misalignment of the RF signals feeding the antenna elements, originated from the random differential delays introduced through the RoF transmission. In this paradigm, the uncompensated skew delays produce a null at $\theta \approx 0°$ and the amplitude of the frequency response is reduced more than 20 dB.

Instead of fully equalizing the delays, the severe radiation pattern distortion can be partially compensated, if we apply phase leads $\varphi_i$ to the antenna elements equal to the residuals modulo $2\pi$, noted as $rem(x, 2\pi)$, of the differential phase lags introduced by the static inter-core skew at the carrier frequency $f_c$, $\varphi_i = rem(2\pi f_c \tau_i, 2\pi)$. The phase leads $\varphi_i$ can be applied to beamforming networks based on phase shifters. In the case of TTD networks, the related time delays for skew compensation are $\tau_i^c = (2\pi - \varphi_i)/2\pi f_c$. These time delays can be applied directly to the OBFN, superimposed on the ordinary delays required to steer the beam to $\theta_0$. This option cancels the requirement for high length accuracy of the full path equalization method, since the optical or electrical beam forming networks already achieve high resolution in the phase shifting process. The compensated transfer function $\widetilde{H_\kappa}$ is

$$\widetilde{H_\kappa} = ab \sum_{i=1}^{N} I_{0,i} e^{-j2\pi(f_c + k/T_s)\tau_i} e^{-j2\pi f_c r_i/c} e^{j\varphi_i} \quad (5)$$

The calculated $P(\theta)$ and $|\widetilde{H_\kappa}|^2$ at $\theta_0=0°$ using (3) and (5) are depicted in fig.2, fig.3 with black lines. The array factor shape is restored, with slightly stronger side-lobes and shallower nulls. This is attributed to the partial compensation process. The phase leads $\varphi_i$ and the related time delays $\tau_i^c$ at the carrier frequency imposed by the OBFN do not entirely equalize the skew delays. The residual skew delays are $\tau_i - \tau_i^c$ and since $\tau_i^c$ are orders of magnitude lower than $\tau_i$, the major part of the delays is left uncompensated. Thus, the antenna feed currents are delayed copies of the RF signal $I_{0,i} \cdot s(t) \cdot e^{j2\pi f_c t}$ (main lobe at $\theta_0=0°$). If those signals are of equal value (uniformly excited Tx antenna), the resultant AF is the one depicted in the blue curve of fig.2. However, as the individual Tx antenna elements are fed with $I_0(i,t) = I_{0,i} \cdot s(t - (\tau_i - \tau_i^c)) \cdot e^{j2\pi f_c t} \approx I_{0,i} \cdot s(t - \tau_i) \cdot e^{j2\pi f_c t}$ they have not the same value for all elements as they depend on $\tau_i$. Consequently, constructive and destructive interference at the receiver antenna of randomly imbalanced electrical fields result in shallower nulls and slightly stronger side lobes (fig. 2, black curve). Again, $|\widetilde{H_\kappa}|^2$ shows a frequency selective behavior as expected. The compensated $|\widetilde{H_\kappa}|^2$ is peaking at the carrier frequency since the phase residuals $\varphi_i$ and associated delays $\tau_i^c$ are calculated to restore the RF phase at this frequency.

The inter-core crosstalk effect on the antenna element currents is also evaluated by setting $x_{ij}$=0.01 (-20 dB). The values of calculated $P(\theta)$ and $|\widetilde{H_\kappa}|^2$ at $\theta_0=0°$ are depicted in fig.2, fig.3 with green lines respectively. The distortion of $P(\theta)$ in fig. 2 of the compensated system is not negligible as we considered an abnormally high value of -20 dB crosstalk for 1 km of MCF fiber. The expected values of optical power crosstalk for such short links are expected to be orders of magnitude lower [14], not posing any noticeable degradation to the antenna radiation pattern and frequency response.

The phase term $2\pi(f_c + k/T_s)\tau_i$ in (5) can be expressed as $2m\pi + rem(2\pi f_c \tau_i, 2\pi) + \frac{2\pi k}{T_s}\tau_i = 2m\pi + \varphi_i + \frac{2\pi k}{T_s}\tau_i$, and so we get $\widetilde{H_\kappa} = ab \sum_{i=1}^{N} I_{0,i} e^{-j2\pi(k/T_s)\tau_i} e^{-j2\pi f_c r_i/c}$. Recalling that $I_{0,i} = |I_{0,i}| e^{-j(i-1)\pi \sin \theta_0}$ in order to compensate the phase offsets due to the path length differences in $r_i$ in the direction $\theta_0$ and $r_i \cong r_0 - y_i \cdot \sin\theta_0$ at far field, $\widetilde{H_\kappa}$ can be expressed as

$$\widetilde{H_\kappa} = ab e^{-j[\frac{2\pi f_c r_0}{c} + \frac{\pi}{2} \cdot \sin\theta_0 \cdot (N-1)]} \sum_{i=1}^{N} |I_{0,i}| e^{-j2\pi(k/T_s)\tau_i} \quad (6)$$

In TTD networks, the time delays $\tau_i^c$ for skew compensation impose slightly different phase leads $\varphi_i^k = 2\pi \left(f_c + \frac{k}{T_s}\right)\tau_i^c = \varphi_i + 2\pi \frac{k}{T_s}\tau_i^c$ in every subcarrier of the OFDM signal. In this case, (6) is expressed as

$$\widetilde{H_\kappa} = ab e^{-j[\frac{2\pi f_c r_0}{c} + \frac{\pi}{2} \cdot \sin\theta_0 \cdot (N-1)]} \sum_{i=1}^{N} |I_{0,i}| e^{-j2\pi(\frac{k}{T_s})(\tau_i - \tau_i^c)} \quad (6a)$$

As already stated, the delays $\tau_i^c$ introduced by the TTD networks are negligible compared to the static skew delays and no differences in frequency response are detected when using (6a) instead of (6). Since the analysis does not encounter the effects of noise, the parameters $a, b$ are only scaling factors and also the phase term $e^{-j[\frac{2\pi f_c r_0}{c} + \frac{\pi}{2} \cdot \sin\theta_0 \cdot (N-1)]}$ in (6) is constant, not affecting the channel frequency response, so both can be considered equal to unity and (6) is reduced to

$$\widetilde{H_\kappa} = \sum_{i=1}^{N} |I_{0,i}| e^{-j2\pi(k/T_s)\tau_i} = \frac{1}{\sqrt{N}} \sum_{i=1}^{N} e^{-j2\pi(k/T_s)\tau_i} \quad (7)$$

It should be noted that the described partial compensation as well as the full compensation schemes require the knowledge of the skew characteristics of the fiber. Therefore, a skew characterization should be done beforehand, preferably including all possible delay mismatches, both at the optical and electrical domain.

IV. STATISTICAL EVALUATION OF STATIC SKEW EFFECTS AND COMPENSATION PROCESSES

Based on the paradigm presented in section III we conducted statistical analysis of $|\widetilde{H_\kappa}|^2$ at $\theta_0$ and $P(\theta)$ for 1000 runs of $N$=8 uniformly excited antenna elements. The inter-core differential delays are samples of uniform distributions with maximum values $\tau_{max}^u$ of 1 ns and 2 ns. In fig. 4, the calculated mean values of $|\widetilde{H_\kappa}|^2$ and $P(\theta)$ are depicted. In fig. 4b, for $\tau_{max}^u$ =2 ns, the main lobe gain is slightly reduced, with small enhancement of the side lobes and elevation of the background level compared to fig. 4a, due to higher skew delays, as already discussed in section III.

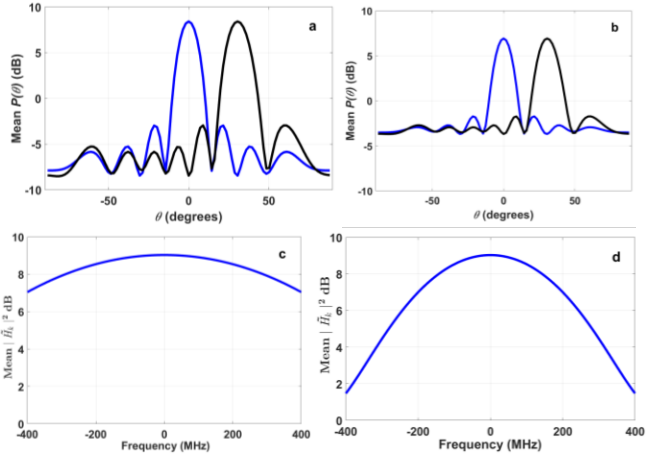

Fig. 4: The statistical mean value of the array factor $P(\theta)$ (a, b) and frequency response $|\widetilde{H_\kappa}|^2$ (c, d) at $\theta_0$ for $N=8$ and $\tau_{max}^u=$1ns (a, c), $\tau_{max}^u=$2ns (b, d). 30º steered beam with uniform current distribution is depicted in black.

Moreover, a steered lobe to 30º generated by feeding the antenna elements with proper phase is also depicted in figs. 4a,b so as to validate and illustrate the proposed skew compensation method at a different pointing angle of the main lobe. In fig. 4d ($\tau_{max}^u$ =2ns), the frequency response is significantly affected with almost 6 dB decrease between the carrier and ±400MHz frequencies, while for $\tau_{max}^u$ =1 ns (fig. 4c) the decrease is only 2 dB. This is attributed to the minima of the mean $|\widetilde{H_\kappa}|^2$ located at certain frequency offsets from the carrier. The frequencies of the minima are closer to the carrier in the case of 2 ns maximum delay, causing steeper descend of the frequency response in fig. 4(d).

A statistical measure of the "flatness" of $|\widetilde{H_\kappa}|^2$ depending on the delays is to calculate the complementary cumulative distribution functions (ccdfs) for -1 dB power bandwidth, namely $|\widetilde{H_\kappa}|^2 \geq 0.794 \cdot |\widetilde{H_0}|^2$, where $|\widetilde{H_0}|^2$ is the maximum value at the carrier frequency ($k=0$). The resulting ccdfs are depicted in fig. 5a. For $\tau_{max}^u = 1$ ns, the guaranteed -1 dB bandwidth is in the order of 400 MHz, while for 2 ns is reduced to roughly 200 MHz.

## V.  FREQUENCY RESPONSE AND RMS DELAY

In addition to the statistical analysis presented in section IV, where all the skew values were obtained from uniform distributions, we acquired the relation between -1 dB and -3 dB mean bandwidths and the rms delays for uniform, Gaussian, Rayleigh and negative exponential probability density functions of the skew delays. The -1 dB ($B_{-1dB}$) and -3 dB ($B_{-3dB}$) power bandwidths are defined as the bandwidths in which the received power $|\widetilde{H_\kappa}|^2$ is dropping to -1 dB or -3 dB respectively of the maximum $|\widetilde{H_0}|^2$ value. We generated 1000 sets of $N$ delay samples from each distribution, with the same rms delay. The rms delay of the uniform distribution with zero minimum value is calculated from $\tau_{max}^u$ using the formula $\tau_{rms} = (\sqrt{3}/6)\tau_{max}^u$. The samples of the three other distributions are generated with the same $\tau_{rms}$ of the uniform distribution. It should be noted that the mean value of the delay samples is not important as the frequency response and antenna pattern distortion are affected by the differential delays between the MCF cores.

Applying the generated delay values to (7) we obtained results for $N=$4, 8, 16, 32, 64 and 128 antenna elements. It should be noted that available MCFs incorporate as much as 39 cores [15] but larger number of cores is expected in the future. In fig. 5b the calculated minimum, mean and maximum values of -1dB bandwidth for uniform delay distribution of $N=$8 antenna elements are depicted. The mean -1 dB bandwidth for $\tau_{rms}=$0.29 ns, which corresponds to 1 ns of maximum delay, is roughly 600 MHz which is equal to the one extracted from fig. 4c. Similarly, the minimum value (black curve) of $B_{-1dB}$ is 390 MHz for $\tau_{rms}=$0.29 ns, roughly equal to the guaranteed -1 dB bandwidth obtained from the ccdf for $\tau_{max}^u=$ 1ns in fig. 5a. We also approximated the obtained bandwidth values with respect to the rms delay with the inverse proportion function $B_{-1dB} = D/\tau_{rms}$ [16]. The values of $D$ are 114.4 MHz·ns, 176.2 MHz·ns and 479.0 MHz·ns for the minimum, mean and maximum $B_{-1dB}$ values respectively. In table I, the values of $D$ for mean $B_{-1dB}/B_{-3dB}$ bandwidths with respect to the delay distributions and number of antenna elements are presented. The calculated values with uniform, Gaussian and Rayleigh distributed delays converge to 156 MHz·ns for $B_{-1dB}$ and 284 MHz·ns for $B_{-3dB}$ respectively, for large number ($N=$128) of antenna elements. Slightly wider bandwidths are achieved when the delays follow the negative exponential distribution. The same behavior is observed for the minimum and maximum values of $B_{-1dB}, B_{-3dB}$ bandwidths. Typical converged values of $D$ are

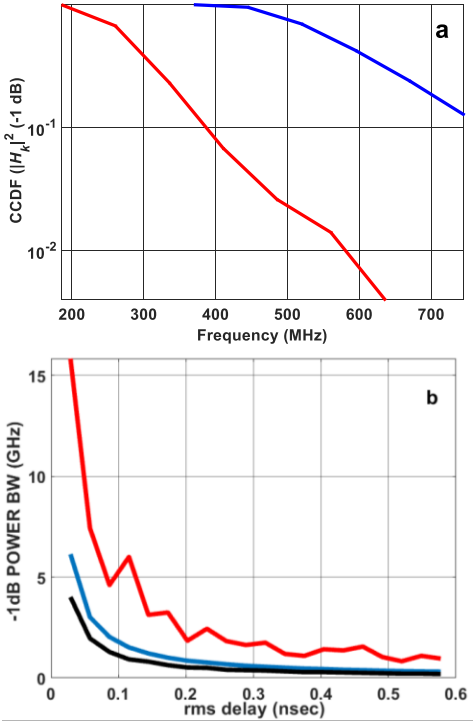

Fig. 5: (a), the ccdfs of $|\widetilde{H_\kappa}|^2 \geq 0.794 \cdot max(|\widetilde{H_\kappa}|^2)$ for $N=$8 elements, red for $\tau_{max}^u = $ 2ns and blue for $\tau_{max}^u = $ 1ns. (b), -1dB minimum (black), mean (blue) and maximum (red) power bandwidth of $|\widetilde{H_\kappa}|^2$ for 1000 sets of $N=$8 elements, uniform distribution of delays.

140 MHz·ns minimum, 190 MHz·ns maximum for $B_{-1dB}$ and 220 MHz·ns minimum, 340 MHz·ns maximum for $B_{-3dB}$.

As stated in section II, there is no statistical evaluation of the skew delays available. Nevertheless, we can reasonably assume that the Gaussian distribution describes more accurately the skew statistics for large number of samples. On the other hand, the negative exponential distribution seems to underestimate the skew effects with respect to all other distributions (table I). An important result of our analysis is that irrespective of the skew delay distribution, the mean bandwidth - rms delay product, $D$, converges to the same values for large antenna arrays.

TABLE I

VALUES OF $D$ (MHz·ns) FOR MEAN $B_{-1dB}/B_{-3dB}$

| $N$ | Uniform | Gaussian | Rayleigh | Neg. exp. |
|-----|---------|----------|----------|-----------|
| 4   | 219.2/370.6 | 239.4/405.6 | 242.0/410.0 | 316.0/535.6 |
| 8   | 176.2/299.6 | 184.2/317.0 | 186.0/319.4 | 222.6/402.4 |
| 16  | 165.6/280.4 | 166.0/287.0 | 167.2/289.8 | 186.4/362.8 |
| 32  | 161.0/272.0 | 159.6/275.4 | 160.4/278.4 | 175.0/341.8 |
| 64  | 159.2/268.8 | 156.2/269.6 | 157/272.4 | 167.8/328.8 |
| 128 | 158.0/266.8 | 155.0/267.6 | 155.8/270.6 | 165.6/323.8 |

## VI. NUMERICAL RESULTS

The theoretical results presented in section III are validated through simulations of a MCF – ARoF and radio link, using the differential delays of the MCF cores. The ARoF transmitter comprises the OFDM baseband transmitter and a 26 GHz up-converter along with a typical 1550 nm laser and external Mach Zehnder (MZM) modulator for optical transmission. The optical beamforming network is simulated as a TTD device with one input and $N=4$ outputs. The MCF is considered as parallel SMF channels with negligible crosstalk and adjustable sample delays which simulate the skew. The fiber transmission is simulated by solving the non-linear Schrödinger equation using the split step Fourier method. Dispersion and non-linear effects are taken into account as well. The ARoF receiver includes $N=4$ parallel photodiode – transimpedance amplifier (TIA) and RF Power Amplifier chains which feed the 4 element ULA, as shown in fig. 1. Free space path loss and individual phase progression per transmitter antenna element is considered during the radio wave propagation to the receiver antenna. The receiver consists of the RF down-conversion module as well as the channel estimation and OFDM demodulation subsystems. The channel estimation process is accomplished by inserting pilot signals for every 8 subcarriers. The complete transfer function is acquired with interpolation on the data subcarriers. To attain simulation validity and completeness, typical noise sources are inserted in all relevant stages of optical and RF subsystems. Then, noise contribution is washed out by averaging the numerical results.

The simulation model parameters are presented in table II. The numerical results depicted in fig. 6 for the array factor $P(\theta)$ (markers) are obtained by estimating the average received power of 4 successive OFDM symbols per angle, at different angles relative to the transmitter antenna.

TABLE II

SIMULATION MODEL PARAMETERS FOR THE AROR/RF LINK

| Parameter | Value |
|-----------|-------|
| Modulation per subcarrier | 16-QAM |
| OFDM sampling frequency | 983.04 MHz |
| FFT/IFFT size $N_F$ | 4096 |
| Number of used subcarriers $N_{Fu}$ | 3332 |
| OFDM symbol duration $Ts$ (w/o CP) | 4.16us |
| Subcarrier spacing, $1/Ts$ | 240KHz |
| Cyclic Prefix (CP) | 520ns |
| Total symbol duration | 4.68 us |
| Signal bandwidth | 800 MHz |
| Carrier frequency $fc$ | 26 GHz |
| Model sampling frequency | 251.66 GHz |
| Laser linewidth | 600 kHz |
| MZM max driving signal p-p | $0.7 \cdot V_\pi$ |
| OBFN steering angle $\theta_0$ | 0° |
| Launched optical power per core | 0 dBm |
| Fiber dispersion | $-20 \cdot 10^{-27}$ s$^2$/m |
| Fiber non-linear parameter | $1.2 \cdot 10^{-3}$ (W·m)$^{-1}$ |
| Fiber attenuation | 0.2 dB/km |
| MCF length | 1 km |
| MCF skew delays, $\tau_i$ | As in section III |
| RF PA output per element | +15 dBm |
| Tx antenna elements, $N$ | 4 |
| Radio transmission path length | 100 m |
| Receiver antenna gain | 0 dBi |
| Receiver gain | 30 dB |
| Receiver noise figure | 8 dB |

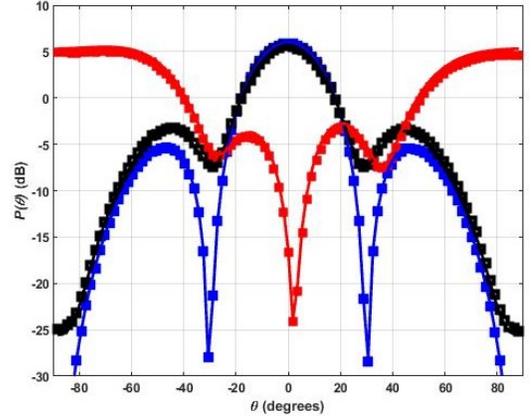

Fig. 6: Theoretical (solid lines) vs numerical (markers) results. Blue: w/o skew, red: w/ skew, black: w/ compensation.

The OBFN at the transmitter is adjusted to provide equal phase to the signals entering all MCF cores if no compensation is performed (blue and red markers in fig. 6). As a result, maximum antenna gain is observed (blue markers) at $\theta_0 = 0°$, broadside to the array, when no skew effect is considered. The differential delays introduced by the MCF transmission severely distort the radiation pattern (red markers). When applying the proposed partial compensation procedure to the OBFN, the radiation pattern is restored (black markers). The array factor curves are very well fitted to the theoretical ones (solid lines) in all cases as shown in fig. 6. It is observed that by using typical values for the parameters of the ARoF link, the consistency of the numerical and theoretical results is realized, since MZM nonlinearities are minimal as the peak to peak drive voltage is $0.7 \cdot V_\pi$, fiber induced nonlinearities are negligible at 0 dBm power levels and spatial fading due to fiber dispersion is not affecting the considered 1 km link at 26 GHz.

Moreover, the frequency response $|\widetilde{H_\kappa}|^2$ at $\theta_0 =0°$ is obtained after averaging 50 simulation runs and depicted in fig. 7. We used root raised cosine filters both at the transmitter and receiver. The half bandwidth frequency response of the filter is depicted in red color in fig. 7. The simulation results (blue) are following the filter response (red) and at the same time undergo the low pass behavior of the theoretical prediction (black). The -1 dB filter response valley at 170 MHz and +0.3 dB peak at 325 MHz are imprinted to the numerically obtained curve, taking into account both the transmitter and receiver filter contribution together with the theoretically obtained 0.2 dB and 0.8 dB attenuation due to the skew effect. Hence it is seen that the numerical results are in perfect agreement with the theoretical ones, thus validating the analytical model and proposed partial skew compensation method.

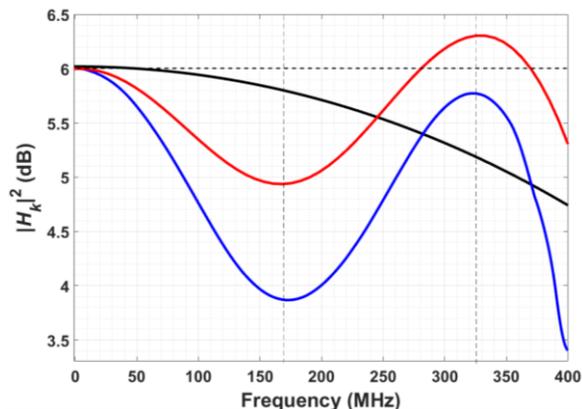

Fig. 7: Half bandwidth frequency response for $\theta_0 =0°$. Black: Theoretical w/compensation, red: root raised cosine filter response, blue: simulated w/compensation.

## VII. CONCLUSION

We performed a theoretical analysis for quantifying the MCF skew effects and introduced an efficient method for MCF static skew compensation in analog RoF beamforming. The effects of the residual uncompensated skew on the frequency response and the array factor are calculated. The theoretical analysis and predictions are supported and validated with simulations. The obtained results can be used in the characterization and performance evaluation of remote analog beamforming systems. The knowledge of the statistical parameters, namely the rms value of the differential delays, is sufficient to estimate the bandwidth of the partially compensated system with low uncertainty, especially for large antenna arrays. Further work focuses not only on the experimental evaluation of the proposed method, but also on the investigation of an automatic procedure for acquiring the static skew delays remotely, thus not requiring the prior to installation knowledge of this quantity.

REFERENCES

[1] M. Shafi, A. F. Molisch, P. J. Smith, T. Haustein, P. Zhu, P. D. Silva, & G. Wunder, "5G: A tutorial overview of standards, trials, challenges, deployment, and practice", *Journal on Selected Areas in Communications*, vol. 35, no. 6, pp. 1201-1221, 2017.

[2] L. Wei, R. Q. Hu, Y. Qian, & G. Wu, "Key elements to enable millimeter wave communications for 5G wireless systems", *IEEE Wireless Communications*, vol. 21, no. 6, pp. 136-143, 2014.

[3] S. E. Alavi, M. R. K. Soltanian, I. S. Amiri, M. Khalily, A. S. M. Supa'at, & H. Ahmad, "Towards 5G: A photonic based millimeter wave signal generation for applying in 5G access fronthaul", *Scientific reports*, *6*, 19891, June 2016.

[4] G. K. M. Hasanuzzaman, & S. Iezekiel, "Multi-core Fiber Based Mm-wave Generation, Radio-over-Fiber, and Power-over-Fiber", *IEEE 2018 11th International Symposium on Communication Systems, Networks & Digital Signal Processing (CSNDSP)*, pp. 1-3, July 2018.

[5] T. Nagayama, S. Akiba, T. Tomura, & J. Hirokawa, "Photonics-Based Millimeter-Wave Band Remote Beamforming of Array-Antenna Integrated with Photodiode Using Variable Optical Delay Line and Attenuator", *Journal of Lightwave Technology*, vol. 36, no. 19, pp. 4416-4422, October 2018.

[6] L. Zhuang, C. G. H. Roeloffzen, R. G. Heideman, A. Borreman, A. Meijerink, & W. van Etten, "Single-Chip Ring Resonator-Based 1X8 Optical Beam Forming Network in CMOS-Compatible Waveguide Technology". *IEEE Photonics Technology Letters*, vol. 19, no. 15, pp. 1130-1132, August 2007.

[7] B. J. Puttnam, G. Rademacher, R. S. Luís, J. Sakaguchi, Y. Awaji, & N. Wada, "Inter-Core Skew Measurements in Temperature Controlled Multi-Core Fiber", *IEEE 2018 Optical Fiber Communications Conference and Exposition* (*OFC*), pp. 1-3, March 2018.

[8] T. Umezawa, P. T. Dat, K. Kashima, A. Kanno, N. Yamamoto, & T. Kawanishi, "100-GHz Radio and Power Over Fiber Transmission Through Multicore Fiber Using Optical-to-Radio Converter", *Journal of Lightwave Technology*, vol. 36*, no. 2, pp. 617-623, January 2018.

[9] B. J. Puttnam et al., "High capacity transmission systems using homogeneous multi-core fibers," J. Lightw. Technol., vol. 35, no. 6, pp. 1157–1167, Mar. 2017.

[10] T. Nikas, E. Pikasis, & D. Syvridis, "Static Skew Compensation in Multi Core Radio over Fiber systems for 5G Mmwave Beamforming," in *2018 Photonics in Switching and Computing (PSC)* (pp. 1-3), Cyprus, September 2018. IEEE.

[11] H. Bach, "Directivity of basic linear arrays", *IEEE Transactions on Antennas and Propagation*, 18(1), 107-110, 1970.

[12] C. A. Balanis, "Antenna theory: A review", *Proceedings of the IEEE*, 80(1), 7-23, 1992.

[13] M. A. Uman, D. K. McLain, & E. P. Krider, "The electromagnetic radiation from a finite antenna", *American Journal of Physics*, *43*(1), 33-38, 1975.

[14] T., Hayashi, T. Taru, O. Shimakawa, T. Sasaki, & E. Sasaoka, "Design and fabrication of ultra-low crosstalk and low-loss multi-core fiber", *Optics express*, vol. 19, no. 17, pp. 16576-16592, August 2011.

[15] Y. Sasaki et al., "Single-mode 37-core fiber with a cladding diameter of 248 μm," in Proc. Opt. Fiber Commun. Conf. Exhib., Los Angeles, CA, USA, Mar. 2017, Paper Th1H.2.

[16] M. Tlich, G. Avril, & A. Zeddam, "Coherence Bandwidth and its Relationship with the RMS delay spread for PLC channels using Measurements up to 100 MHz", *Home Networking* pp. 129-142, Springer, Boston, MA, 2008.